\documentclass{elsart}
\usepackage{epsfig}
\usepackage{amssymb}

\begin{document}

\begin{frontmatter}

\title{Electronic Polarization in Quasilinear Chains}

\author{Michael Springborg,$^{a1}$ \,Bernard Kirtman,$^{b2}$\, and Yi Dong$^{a3}$}
\thanks[label1]{e-mail: m.springborg@mx.uni-saarland.de}
\thanks[label2]{e-mail: kirtman@chem.ucsb.edu}
\thanks[label3]{e-mail: y.dong@mx.uni-saarland.de}

\address{$^a$ Physical Chemistry, University of Saarland, 66123 Saarbr\"ucken, Germany
\\ $^b$ Department of Chemistry and Biochemistry, University of California, Santa Barbara,
California 93106, U.S.A.}

\date{\today }

\begin{abstract}

Starting with a finite $k$-mesh version of a well-known equation by Blount, we 
show how various definitions proposed for the polarization of long chains are related.
Expressions used for infinite periodic chains in the 'modern theory of polarization' 
are thereby obtained along with a new single particle formulation. Separate 
intracellular and intercellular contributions to the polarization are identified and,
in application to infinite chains, the traditional sawtooth definition is found to be
missing the latter. For a finite open chain the dipole moment depends upon how the
chain is terminated, but the intracellular and intercellular polarization do not. All 
of these results are illustrated through calculations with a simple H\"uckel-like
model.

\end{abstract}
\end{frontmatter}

The purpose of this Letter is to answer some questions that arise in 
connection with the theoretical treatment of macroscopic polarization 
in quasi-one dimensional chains. In order to specify the issues let us 
consider a macroscopic, open-ended, polymeric chain consisting of identical 
unit cells. For sake of argument this chain is assumed to be polarized due 
to an asymmetric unit cell and/or an external electric field. There are 
two contributions to the polarization $P$, i.e. to the dipole moment per unit 
cell. One is due to the asymmetric charge distribution within a unit 
cell in the central region of the chain and the other is due to the charge 
of opposite sign that accumulates at the chain ends. H\"uckel-type calculations 
(see below) show that the contribution due to the finite chain ends does not 
vanish even in the infinite chain limit. This is simply due to the fact that, 
for a charge of fixed magnitude at either end, the dipole moment is directly 
proportional to the distance between the charges.

Next, imagine that the chain ends are connected to form a 
ring. In that event there are no ends and all the unit cells are identical. 
What is the relationship between the unit cell charge distribution of the 
closed chain and the unit cell charge distribution at the center of the 
open chain? In fact, they are the same as our H\"uckel-type calculations 
confirm. Then, what has happened to the contribution to the polarization 
associated with the charge build-up at the ends of the open chain? 
As it turns out this contribution is associated with a charge flow term 
that arises from Blount's theoretical expression \cite{blount} for the 
polarization when periodic boundary conditions are applied. This raises 
the question of whether or not such a term can be accounted for by the conventional
sawtooth approach \cite{kr,otto}. The latter is based on using a 
finite mesh in $k$-space, along with periodic boundary conditions \cite{resta,pccp},
but it does not correspond to a finite-mesh analogue of Blount's
formula which will be presented here. From a general formulation of this
analogue several approximations will be developed including 
the fundamental equation(s) of the so-called 
modern theory of polarization \cite{resta1,resta3,ksv,vks,ng}

Using a H\"uckel model, discussed below, we demonstrate quantitatively
that the sawtooth approach omits
the current term and that Blount's formula gives an accurate 
approximation for the polarization when applied with a finite set 
of $k$ points. 
Finally, it will be seen that the polarization of 
long finite chains with arbitrary terminal substituents does not depend 
on the nature of these substituents even though the same cannot be said of 
the dipole moment $D$ itself.

\begin{figure}
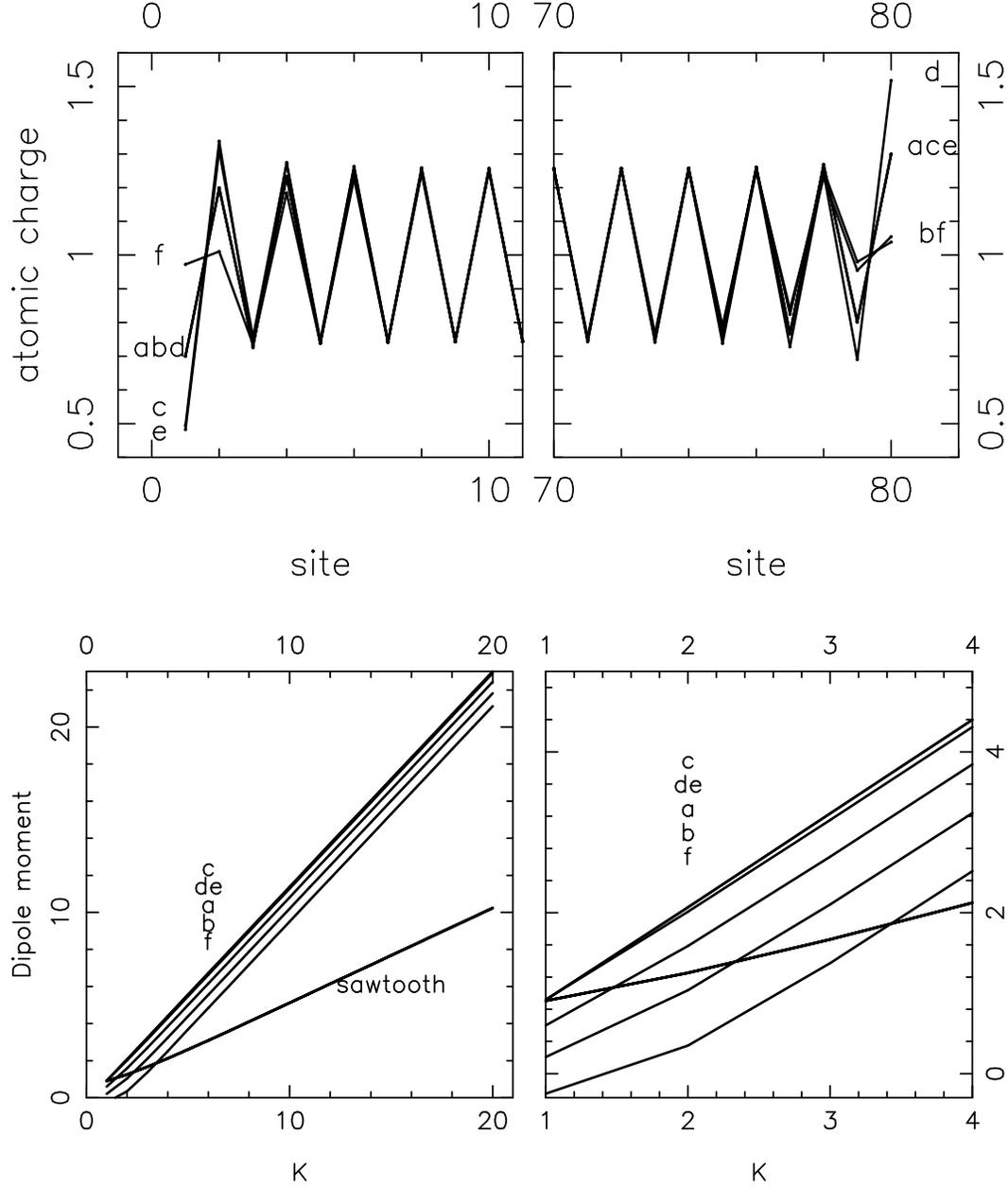

\begin{picture}(14,16)
\put(0,8.5){\psfig{file=xfinrho.ps,width=14cm}}
\put(0,0){\psfig{file=xfinpol.ps,width=14cm}}
\end{picture}
\caption{Upper part: Distribution of the atomic charges as a function of atom index for
a finite chain with 80 atoms for different cases of the matrix elements for the 
terminating atoms, i.e., the on-site energies for the first (last) atom have
been modified as $\epsilon_0\to\epsilon_0+\Delta\epsilon_l$ ($-\epsilon_0\to
-\epsilon_0+\Delta\epsilon_r$), and the 
hopping integrals between the first (last) two atoms according to $-t_+\to -t_++\Delta t_l$ 
($-t_+\to -t_++\Delta t_r$). 
Lower part: The dipole moment for the same cases but as a
function of chain length. Here, `sawtooth' corresponds to 
the dipole moment for a ring system when using the sawtooth approximation, and the
right panel shows a blow-up of the low-$K$ part. The curves
marked $a$, $b$, $c$, $d$, $e$, and $f$ (in the lower panels these labels are listed
in the same order as the curves appear) correspond to the following modifications:
$(\Delta\epsilon_l,\Delta\epsilon_r,\Delta t_l,\Delta t_r)=$ $(0,0,0,0)$, $(1.0,0,0,0)$,
$(0,1.0,0,0)$, $(0,0,-1.0,0)$, $(0,0,0,-1.0)$, and $(1.0,-1.0,-1.0,1.0)$, 
respectively.}
\label{fig01}
\end{figure}

For a finite chain of alternating A and B atoms the H\"uckel Hamiltonian may be 
written in terms of orthonormal atom-centered basis functions $\{\chi_p\}$ as 
\begin{equation}
\hat H=\sum_{p=-2K+1}^{2K}\epsilon_p\hat c_p^\dag\hat c_p-\sum_{p=-2K}^{2K-1}
t_{p,p+1}(\hat c_{p+1}^\dag\hat c_p+\hat c_p^\dag\hat c_{p+1})
\label{eqnx01}
\end{equation}
where $4K$ is the number of atoms, $\hat c_p^\dag$ and $\hat c_p$ are
the creation and annihilation operators for the function $\chi_p$, and $\epsilon_p$
and $-t_{p,p+1}$ are on-site energies and hopping integrals, respectively.
The matrix elements of the position operator are given 
by $\langle \chi_n\vert \hat z\vert\chi_m\rangle=\delta_{n,m}z_n$ with $z_p$ being the 
position of the $p$th atom. For convenience, the atoms are taken to be equally 
spaced, $z_p=\frac{a}{2}\big[p-\frac{1}{2}\big]$, with $p=-2K+1,\dots,2K$
being odd (even) for the A (B) atoms, and $\frac{a}{2}$ being the 
nearest-neighbor distance. Assuming one electron per atom it is straightforward
to evaluate the atomic charges and the electronic polarization, $P$ (dipole
moment per A-B unit) as a function of the number of A-B units, $2K$.
As an example (using arbitrary units for length and energy and setting
the electronic charge equal to $+1$) for $a=2.0$, 
alternating on-site energies $\epsilon_p=\pm 0.5\equiv\pm\epsilon_0$, and hopping
integrals $t_+=-2.2$, $t_-=-1.8$, we find that
the polarization is converged to a value $P = 0.58125$ for $K>13$  while the charges
on the central atoms are $Q({\rm A})=0.74413$, $Q({\rm B})=1.25587$. The latter 
result in a contribution $P_{\rm c}=P-\frac{a}{4}[Q({\rm B})-Q({\rm A})]=0.32538$
to the polarization due to accumulation of charge at 
the chain ends. We may think of $P_{\rm c}$ as an intercellular charge flow term. 
The existence of a substantial intercellular charge flow term is 
remarkably robust to variations in the model. Thus, including next-nearest-neighbor
interactions, modifying the matrix elements at the chain ends 
or altering $\langle \chi_n\vert \hat z\vert\chi_m\rangle$ in 
realistic ways often changes the total dipole moment, but not the 
polarization, of sufficiently long chains. The effect of varying 
terminal (on-site and/or hopping) matrix elements is 
illustrated in Fig.\ \ref{fig01}. The figure shows that altering
the chain ends changes the charges on the ends (upper part) and the dipole 
moment (lower part) but neither the polarization (slope of dipole-moment curve)
nor the charges in the central region are affected. The right hand panel
of the lower part shows that the polarization can vary for small chains
(see, particularly, curve f). 

The fact that one may calculate the polarization by 
studying only the central cells has been shown previously by Vanderbilt 
and King-Smith \cite{vks}. However, this does not imply that different 
terminating groups, which lead to different charge accumulation
at the chain ends, give the same polarization because there is a contribution
due to intercellular charge flow that could change. Thus, this is a generalization 
of the Vanderbilt and King-Smith result, which is consistent with the known
near-sightedness \cite{wk} of the single-particle density matrix. It has obvious 
implications for the design of donor-acceptor, or push-pull, systems and is
valid provided the chain is sufficiently long.

Even for an unsubstituted chain where the ends are connected so that no
charge can accumulate there is an important contribution to the polarization that
arises from the intercellular charge flow. We now turn to that case and consider
a ring of $2K$ identical AB unit cells. Application of periodic boundary conditions
leads to the general expression for the eigenfunctions  
\begin{equation}
\psi_n^k(\vec r)=u_n^k(\vec r) e^{ikz}=\frac{1}{\sqrt{2K}}\sum_{m=-K+1}^{K} e^{ikam} 
\sum_{p=1}^{N_b} c_{pn}^k\chi_{pm}(\vec r),
\label{eqnx02}
\end{equation}
with $n$ being a band index, $\chi_{pm}$ the $p$th basis function of the $m$th 
unit cell, and $N_b$ the number of basis functions per cell. 
In our H\"uckel model, $n=1$ ($n=2$) 
for the occupied (empty) band, and $p=1$ ($p=2$) indicates the function on 
the A (B) atom. For any given set of parameters and sufficiently long chains 
the electronic charges, 
$Q({\rm A})$ and $Q({\rm B})$, turn out to be identical to those at the center of the 
open-ended chain of the same length. Thus, for either chain, the same intracellular 
polarization $\frac{a}{4}[Q({\rm B})-Q({\rm A})]$ is obtained. This means that the 
$P_{\rm c}$ contribution must be accounted for in some other way. For the ring
we can readily identify that contribution by considering Blount's expression for 
the polarization in the limit $K\to\infty$:
\begin{equation}
P=\frac{ia}{\pi}\sum_n\int \langle u_n^k\vert \frac{\partial}{\partial k}u_n^k\rangle dk,
\label{eqnx03}
\end{equation}
where the $n$ summation is over the (doubly) occupied bands.
Using Eq.\ (\ref{eqnx02}) it is easy to show (cf.\ \cite{kgb}) that 
\begin{eqnarray}
P=\frac{1}{\pi}\sum_n\sum_{m=-N}^N \int e^{ikma} \sum_{pq} c_{qn}^{k*}\bigg(\langle \chi_{q0}\vert
z-ma\vert\chi_{pm}\rangle +i\langle \chi_{q0}\vert\chi_{pm}\rangle \frac{d}{dk}\bigg)c_{pn}^k dk.
\nonumber\\
\label{eqnx05}
\end{eqnarray}
With a finite $2K$-point-mesh approximation for the integral one can 
verify that the first term on the rhs of Eq.\ (\ref{eqnx05}) yields the 
intracellular polarization. This leaves the second term as the periodic 
cyclic chain analogue of the intercellular charge flow contribution 
described above in connection with the open chain.   

If one is interested in long open-ended chains it is usually 
advantageous computationally to assume that the chain is infinite and periodic.
A number of different proposals have been advanced for calculating the 
polarization of infinite periodic chains using finite $k$ mesh methods. In order
to compare these approaches we follow the treatment of Blount \cite{blount}, based
on the relation 
\begin{equation}
z\psi_n^k(\vec r)=i e^{ikz}\frac{\partial}{\partial k}e^{-ikz}\psi_n^k(\vec r)
-i\frac{\partial}{\partial k}\psi_n^k(\vec r),
\label{eqnx06}
\end{equation}
to obtain the effect of the coordinate $z$ acting on 
a single electron whose orbital, $\psi(\vec r)$, is expanded
in terms of Bloch waves 
\begin{equation}
\psi(\vec r)=\sum_k \sum_n \psi_n^k(\vec r) f_n^k=\sum_k \sum_n e^{ikz}u_n^k(\vec r) f_n^k.
\label{eqn01}
\end{equation} 
In fact, Blount \cite{blount} obtained Eq.\ (\ref{eqnx05}) by using Eq.\ (\ref{eqnx06}) 
on $\psi$ of Eq.\ (\ref{eqn01}). 
Here instead of a continuous $k$ we will use a finite $k$-mesh, which corresponds to 
assuming that the system possesses the periodicity of the Born von K\'arm\'an (BvK) zone 
containing $2K$ unit cells. Consequently, the analytical derivatives in Blount's 
formulation will be replaced by numerical derivatives. In lowest order the numerical 
derivatives corresponding to the terms in Eq.\ (\ref{eqnx06}) are:
\begin{eqnarray}
\hat\Delta_-' \psi(\vec r)&=&\frac{1}{\Delta k}\sum_k\sum_n
\bigg[\psi_n^k(\vec r) f_n^k-\psi_n^{k-1}(\vec r) f_n^{k-1}\bigg]\nonumber\\
\hat\Delta_-'' \psi(\vec r)&=&\frac{1}{\Delta k}\sum_k\sum_n e^{ikz}
\bigg[u_n^k(\vec r) f_n^k-u_n^{k-1}(\vec r) f_n^{k-1}\bigg]\nonumber\\
\hat\Delta_+' \psi(\vec r)&=& \frac{1}{\Delta k}\sum_k\sum_n
\bigg[\psi_n^{k+1}(\vec r) f_n^{k+1}-\psi_n^k(\vec r) f_n^k\bigg]\nonumber\\
\hat\Delta_+'' \psi(\vec r)&=&\frac{1}{\Delta k}\sum_k\sum_n e^{ikz}
\bigg[u_n^{k+1}(\vec r) f_n^{k+1}-u_n^k(\vec r) f_n^k\bigg]\nonumber\\
\hat \Delta_0'&=&\frac{1}{2}\big(\hat\Delta_-'+\hat\Delta_+'\big)\nonumber\\
\hat \Delta_0''&=&\frac{1}{2}\big(\hat\Delta_-''+\hat\Delta_+''\big),
\label{eqn02}
\end{eqnarray}
with $\Delta k = \frac{\pi}{aK}$. By construction these expressions have
the BvK periodicity and it follows that the lowest order 
finite-$k$-mesh analogues of $\langle\psi\vert z\vert\psi
\rangle$ are $(\psi_n^{k+2\pi/a} = \psi_n^k)$: 
\begin{eqnarray}
\langle \psi\vert (-i\hat\Delta_-'+i\hat\Delta_-'')\vert \psi\rangle&=&
\langle \psi\vert \frac{i}{\Delta k}(1-e^{i\Delta k z})\vert \psi\rangle
=\frac{i}{\Delta k}(1-S^+)\nonumber\\
\langle \psi\vert (-i\hat\Delta_+'+i\hat\Delta_+'')\vert \psi\rangle&=&
\langle \psi\vert \frac{i}{\Delta k}(e^{-i\Delta k z}-1)\vert \psi\rangle
=\frac{i}{\Delta k}(S^--1)\nonumber\\
\langle \psi\vert (-i\hat\Delta_0'+i\hat\Delta_0'')\vert \psi\rangle&=&
\langle \psi\vert \frac{\sin(\Delta k z)}{\Delta k}\vert \psi\rangle
=\frac{1}{2i\Delta k}(S^+-S^-),
\label{eqn03}
\end{eqnarray}
where $S^\pm=\langle\psi\vert e^{\pm i\Delta k z}\vert\psi\rangle$. 
If the spatial extent of $\psi$ is much smaller than $\frac{1}{\Delta k}$ (this can,
e.g., be obtained by increasing the number of $k$ points in an actual calculation),
and assuming that $\int_{\rm BvK} \vert\hat z\psi(\vec r)\vert^2 d\vec r$
exists (e.g. when $\psi$ is a well-localized Wannier function),
then we may make the approximation
\begin{equation}
\langle \psi\vert \frac{i}{\Delta k}e^{\pm i\Delta k z}\vert \psi\rangle \simeq 
\frac{i}{\Delta k}e^{\pm i\Delta k \langle \psi \vert z\vert \psi\rangle}
\label{eqn05}
\end{equation}
or
\begin{eqnarray}
\langle \psi\vert z\vert \psi\rangle &\simeq& \frac{-i}{\Delta k}\ln S^+
\simeq \frac{-1}{\Delta k}{\rm Im} \ln S^+\nonumber\\
\langle \psi\vert z\vert \psi\rangle &\simeq& \frac{i}{\Delta k}\ln S^-
\simeq \frac{1}{\Delta k}{\rm Im} \ln S^-\nonumber\\
\langle \psi\vert z\vert \psi\rangle &\simeq& \frac{1}{\Delta k}{\rm Arcsin}[\frac{1}{2i}
(S^+-S^-)].
\label{eqn04}
\end{eqnarray}
Here, the second equalities in the first two expressions have been obtained by
removing the imaginary parts and, accordingly, requiring that $\langle\psi\vert 
z\vert\psi\rangle$ is real. This result comes about automatically in
the expressions based on the $\hat\Delta_0$ operators.

The treatment for $N$ electrons is similar. In that case the one-electron
Bloch waves are replaced by Slater determinants
$\Psi_{\vec i}^{\vec k}=\hat{\mathcal A}\big[\psi_{i_1}^{k_1}(\vec r_1)\psi_{i_2}^{k_2}
(\vec r_2)\cdots\psi_{i_N}^{k_N}(\vec r_N)]$, where
$\hat{\mathcal A}$ is the antisymmetrizer, and Eq.\ (\ref{eqnx06}) becomes 
\begin{eqnarray}
\bigg(\sum_{n=1}^N z_n\bigg)\Psi_{\vec i}^{\vec k}&=&i \exp (i\sum_{n=1}^N k_nz_n)
\bigg(\sum_{n=1}^N\frac{\partial}{\partial k_n}\bigg)\bigg(\exp(-i\sum_{n=1}^N k_nz_n)
\Psi_{\vec i}^{\vec k}\bigg)\nonumber\\
& &-i\bigg(\sum_{n=1}^N\frac{\partial}{\partial k_n}\bigg)
\Psi_{\vec i}^{\vec k}.
\label{eqny06}
\end{eqnarray}
Then an arbitrary $N$-electron function can be written as the linear combination
$\Psi(\vec r_1,\vec r_2,\dots,\vec r_N)=\sum_{\vec i}\sum_{\vec k}\Psi_{\vec i}^{\vec k}
(\vec r_1,\vec r_2,\dots,\vec r_N)f_{\vec i}^{\vec k}$ with the single-particle situation
being a special case. 
The generalization of the quantities in Eq.\ (\ref{eqn02}) becomes
\begin{eqnarray}
\hat\Delta_-'\Psi(\vec r_1,\vec r_2,\dots,\vec r_N)&=&\frac{1}{\Delta k}\sum_{\vec i}
\sum_{\vec k}\bigg[
\Psi_{\vec i}^{\vec k}(\vec r_1,\vec r_2,\dots,\vec r_N)f_{\vec i}^{\vec k}\nonumber\\
& & \qquad
-\Psi_{\vec r}^{\vec k-\Delta\vec k}(\vec r_1,\vec r_2,\dots,\vec r_N)
f_{\vec i}^{\vec k-\Delta\vec k}\bigg]\nonumber\\
\hat\Delta_-''\Psi(\vec r_1,\vec r_2,\dots,\vec r_N)&=&\frac{1}{\Delta k}\hat{\mathcal A}\Bigg\{
\sum_{\vec i}\sum_{\vec k}e^{i(k_1z_1+k_2z_2+\cdots k_Nz_N)}\nonumber\\
& &\qquad\times\bigg[ u_{i_1}^{k_1}
(\vec r_1)u_{i_2}^{k_2}(\vec r_2)\cdots u_{i_N}^{k_n}(\vec r_N)f_{\vec i}^{\vec k}
\nonumber\\& &\qquad-
u_{i_1}^{k_1-1}(\vec r_1)u_{i_2}^{k_2-1}(\vec r_2)\cdots u_{i_N}^{k_n-1}
(\vec r_N)f_{\vec i}^{\vec k-\Delta\vec k}\bigg]\Bigg\}
\label{eqn07}
\end{eqnarray}
with analogous expressions for $\hat \Delta_+'$ and $\hat \Delta_+''$,
$\hat \Delta_0'$ and $\hat \Delta_0''$. Hence, the generalization of Eq.\ (\ref{eqn03}) is
\begin{eqnarray}
\hat\Delta_-&=&-i\hat\Delta_-'+i\hat\Delta_-''=\frac{i}{\Delta k}
\bigg[1-e^{i\Delta k(z_1+z_2+\cdots+z_N)}\bigg]\nonumber\\
\hat\Delta_+&=&-i\hat\Delta_+'+i\hat\Delta_+''=\frac{i}{\Delta k}
\bigg[e^{-i\Delta k(z_1+z_2+\cdots+z_N)}-1\bigg]\nonumber\\
\hat\Delta_0&=&-i\hat\Delta_0'+i\hat\Delta_0''=
\frac{1}{\Delta k}\sin\bigg[\Delta k(z_1+z_2+\cdots+z_N)\bigg].
\label{eqn08}
\end{eqnarray}
We will restrict ourselves to the case where there is a finite
gap between occupied and unoccupied bands, assume 
no spin polarization ($N$ is even), and use a single determinant wavefunction 
(Hartree-Fock or Kohn-Sham theory). Then
\begin{eqnarray}
\langle \Psi\vert\hat\Delta_-\vert \Psi\rangle&=&\frac{i}{\Delta k}\bigg[1-
({\rm det}\,\underline{\underline{S}}^+)^2\bigg]\nonumber\\
\langle \Psi\vert\hat\Delta_+\vert \Psi\rangle&=&\frac{i}{\Delta k}\bigg[
({\rm det}\,\underline{\underline{S}}^-)^2-1\bigg]\nonumber\\
\langle \Psi\vert\hat\Delta_0\vert \Psi\rangle&=&\frac{1}{2i\Delta k}
\bigg[({\rm det}\,\underline{\underline{S}}^+)^2-({\rm det}\,\underline{\underline{S}}^-)^2
\bigg],
\label{eqn09}
\end{eqnarray}
where $\underline{\underline{S}}^\pm$ is the $N/2\times N/2$ matrix containing the 
single-particle
matrix elements $S_{(i,k),(j,l)}^{\pm}=\langle\psi_i^k\vert e^{\pm i\Delta k z}\vert\psi_j^l\rangle$.
Assuming localized orbitals we may apply the analogue of Eq.\ (\ref{eqn05}), i.e. 
\begin{equation}
\langle \Psi\vert \frac{i}{\Delta k}e^{\pm i\Delta k (z_1+z_2+\cdots+z_N)}\vert \Psi\rangle 
\simeq 
\frac{i}{\Delta k}e^{\pm i\Delta k \langle \Psi \vert z_1+z_2+\cdots+z_N\vert \Psi\rangle},
\label{eqn05x}
\end{equation}
and either of the first two equations in Eq.\ (\ref{eqn09}), in combination with
Eq.\ (\ref{eqn08}), to arrive at the expression
\begin{equation}
P_{\rm R}=-\frac{a}{\pi}{\rm Im}\,\ln\,{\rm det}\,\underline{\underline{S}}^+
=\frac{a}{\pi}{\rm Im}\,\ln\,{\rm det}\,\underline{\underline{S}}^-,
\label{eqn10}
\end{equation}
for the polarization.
Note that $(\underline{\underline{S}}^+)^\dag=\underline{\underline{S}}^-$. 
An essentially identical formula in terms of Bloch orbitals has been given
by Resta \cite{resta}. We observe that Eq.\ (\ref{eqn10}) is based on a
not too accurate finite-difference approximation to the derivative. A more 
accurate approximation is
\begin{equation}
P_0=\frac{a}{2\pi}{\rm Arcsin}\bigg[\frac{1}{2i}\big(
({\rm det}\,\underline{\underline{S}}^+)^2-({\rm det}\,\underline{\underline{S}}^-)^2
\big)\bigg].
\label{eqn11}
\end{equation}
Despite this $P_{\rm R}$ turns out to be more useful computationally. To see why we write
\begin{equation}
{\rm det}\,\underline{\underline{S}}^\pm = s\pm i t,
\label{eqn11a}
\end{equation}
whereby 
\begin{eqnarray}
P_{\rm R}&=&-\frac{a}{\pi}{\rm Arctan}(\frac{t}{s})\nonumber\\
P_0&=&\frac{a}{2\pi}{\rm Arcsin}(2st).
\label{eqn11b}
\end{eqnarray}
As $K\to\infty$, $s^2+t^2\to 1$, while $\vert s\vert,\vert t\vert<1$ are increasing functions
of $K$. Accordingly, as  our numerical
results below confirm, $P_{\rm R}$ converges faster than $P_0$ as a function of $K$.  
On the other hand, $P_0$ can be valuable analytically; indeed, it motivated our choice 
for the operator defined in Eq. (\ref{eqn14}) below.  

The value of det $\underline{\underline{S}}^\pm$
will not be altered by an arbitrary unitary transformation of the single determinant
orbitals. So, instead of localized orbitals we may use the occupied Bloch waves from 
which these orbitals are obtained. Then, the matrix elements 
of $\underline{\underline{S}}^\pm$ 
are non-zero only for pairs of Bloch waves whose $k$ values differ by $\Delta k$ 
(modulus $\frac{2\pi}{a}$). As a result $\underline{\underline{S}}^\pm$ can 
be written as consisting of $2K\times 2K$ square blocks, each of dimension 
$B=\frac{N}{4K}$ (the number of doubly-occupied bands) with non-zero elements only
in the set of blocks lying one stripe above and one stripe below the main diagonal.  

Given that the Bloch functions are differentiable with 
respect to $k$ as discussed by Blount we obtain for small $\Delta k$ 
\begin{equation}
\ln ({\rm det}\,\underline{\underline{S}}^\pm)^2\simeq \mp 2\Delta k \sum_{k=1}^{2K}\sum_{n=1}^B
\langle u_n^k\vert \frac{\partial}{\partial k}u_n^k\rangle.
\label{eqn12}
\end{equation}
Inserting this into the rhs of Eq.\ (\ref{eqn10}) 
yields another formula for the polarization
\begin{equation}
P_{\rm KSV}=\frac{i}{K}\sum_{k=1}^{2K}\sum_{n=1}^B \langle u_n^k\vert 
\frac{\partial}{\partial k}u_n^k\rangle,
\label{eqn13}
\end{equation}
which is the 1D discretized Berry phase
expression \cite{vks} used in the modern theory of polarization.
 
For the treatment of core orbitals (or those of non-interacting periodically 
repeated molecules) we suppose that the orbitals are strongly localized so that
$\langle\psi_{p_1}(\vec r-\vec R_{n_1})\vert e^{\pm i\Delta k z}\vert\psi_{p_2}
(\vec r-\vec R_{n_2})\rangle$
vanishes unless the units $n_1$ and $n_2$, where the functions are centered, are identical. 
In that case we may write for orbitals of the same unit
\begin{eqnarray}
\langle\psi_{p_1}\vert e^{\pm i\Delta k z}\vert
\psi_{p_2}\rangle&\simeq&\delta_{p_1,p_2} e^{\pm \Delta k z_{p_1,p_2}}\pm i\Delta k 
\langle\psi_{p_1}\vert z-z_{p_1,p_2}\vert\psi_{p_2}\rangle\nonumber\\
& &+\frac{(i\Delta k)^2}{2}\langle\psi_{p_1}\vert (z-z_{p_1,p_2})^2\vert\psi_{p_2}\rangle+\cdots
\label{eqn13c}
\end{eqnarray}
with $z_{p_1,p_2}$ being the `center' of the $p_1$th and $p_2$th orbital. 
Therefore, the `traditional' 
contribution to the polarization from these orbitals, i.e., $\sum_p \langle\psi_p\vert z\vert
\psi_p\rangle$, is obtained only in the case where all terms but the first one 
on the rhs of Eq.\ (\ref{eqn13c})
are negligible (e.g. in the limit $\Delta k\to 0$). 

So far we have presented an internally consistent approach for how to calculate the 
polarization in an infinite, periodic chain when basing the discussion on a generalization
of Blount's work to the case of a finite BvK zone. We have arrived at an expression involving
the expectation values for $N$-body operators, i.e., the $\underline{\underline{S}}^\pm$
matrices. This has been taken as a proof that the polarization is a many-body 
phenomenon \cite{resta1,resta3}. However, the polarization can also be written 
in terms of the single-particle operator
\begin{eqnarray}
\hat P=\frac{1}{2i\Delta k}\sum_{m=1}^{N}\sum_{k'}\sum_{n'}& &\bigg[e^{i\Delta k z_m}
\vert\psi_{n'}^{k'-1}(\vec r_m)\rangle\langle
\psi_{n'}^{k'}(\vec r_m)\vert\nonumber\\
& & - e^{-i\Delta k z_m}\vert\psi_{n'}^{k'+1}(\vec r_m)\rangle\langle
\psi_{n'}^{k'}(\vec r_m)\vert \bigg].
\label{eqn14}
\end{eqnarray}
It is straightforward to show that the expectation value of this operator 
gives  $P_0$ in the limit $\Delta k\to 0$.

In order to explore our ideas further, the H\"uckel-like model described
above was used to evaluate the various polarization expressions we have
presented. 
For our purposes it is necessary to have matrix elements of
$z$ and  $e^{\pm i\Delta k z}$ that are defined consistently.  
Hence, we calculated the matrix elements in both cases analytically 
assuming piecewise constant basis functions of adjustable width, $w$. For simplicity 
we also assumed that $w<\frac{a}{2}$, whereby the results become independent of $w$. Other more 
realistic functions are possible, of course, but the above choice is sufficient
to make the desired comparisons. 

In Table 1 we show some 
typical results for the various choices of $P$ obtained using BvK periodic boundary conditions.
The finite chain value determined from the increment 
$\Delta D=\frac{1}{2}[D(2K+2)-D(2K)]$, where $D$ is the dipole moment, 
is also presented for comparison.
In order to interpret polarization values the reader should recall that
$P$ is determined only up to an arbitrary multiple of the unit cell length
[cf.\ Eqs.\ (\ref{eqn10}) and (\ref{eqn11})], which
in this case is 2.0. Bearing this in mind, the table shows that 
$\Delta D$ agrees very well with  the polarization of the
infinite system given by $P_{\rm R}$. Indeed, the finite chain result converges more rapidly 
to the infinite $K$ limit. The sawtooth approximation (denoted
$P_{\rm st}$ in the table) is calculated using periodic boundary conditions
with $z$ replaced by a piecewise linear function having the BvK periodicity.
Note that $P_{\rm st}$ gives the correct value only when the system consists of purely 
non-interacting units (last case in table). Since $\sin(\alpha)=
\sin(\pi-\alpha)$ it is not
possible to discriminate between $P_0$ and $\frac{a}{2}-P_0$ (cf.\ the first case
in the table). If that is taken into account, we see that
$P_{\rm KSV}$, $P_{\rm R}$, and $P_0$ all give similar results, although the latter
converges much slower, and the former much faster, than the others.

In conclusion we have provided a unified picture of electronic polarization in extended 
quasilinear chains based primarily on the finite $k$-mesh analogue of  Blount's treatment for 
infinite periodic systems. Separate intracellular and intercellular contributions are 
identified and compared between closed and open chains. It is shown that neither component 
is affected by substitution at the end of an open chain, as occurs in a push-pull compound. 
On the other hand, the traditional sawtooth formulation for infinite closed chains fails to 
account for the intercellular charge flow term. 
Several different expressions for the electronic polarization are systematically generated 
from the same starting point, including those related to the so-called modern theory of 
polarization. From the same perspective we obtain an alternative single particle operator, 
which yields the polarization as its expectation value. H\"uckel-type calculations are 
carried out to illustrate all of these points and to assess the convergence properties of 
the various polarization formulas as the $k$-mesh spacing decreases to zero.

This work was supported by the German Research Council (DFG) 
through project Sp 439/11. One of the authors (MS) is grateful to Fonds der Chemischen
Industrie for generous support.

\clearpage

\begin{table}
\caption{
Results of model calculations with the H\"uckel model. The lattice constant equals $a=2$. 
All other parameter values are given in the table.}
\begin{tabular}{ccccccccc}
$\epsilon_0$ & $t_+$ & $t_-$ & $K$ & $P_{\rm st}$ & $P_{\rm R}$ & 
$P_{\rm KSV}$ & $P_0$ & $\Delta D$ \\
\hline
 0.5 & 2.2 & 1.8 &    20 & 0.25587 & -1.41859 & -1.41745 & 0.29410 & 0.58125\\
     &     &     &   200 & 0.25587 & -1.41875 & -1.41875 & 0.39798 & " \\
     &     &     &  2000 & 0.25587 & -1.41875 & -1.41875 & 0.41643 & " \\
     &     &     & 20000 & 0.25587 & -1.41875 & -1.41875 & 0.41852 & " \\
 0.5 & 2.5 & 1.5 &    20 & 0.21337 & -1.68291 & -1.68305 & 0.26857 & 0.31695\\
     &     &     &   200 & 0.21337 & -1.68305 & -1.68305 & 0.31134 & " \\
     &     &     &  2000 & 0.21337 & -1.68305 & -1.68305 & 0.31638 & " \\
     &     &     & 20000 & 0.21337 & -1.68305 & -1.68305 & 0.31690 & " \\
 0.5 & 1.5 & 1.5 &    20 & 0.33562 & -1.00000 & -3.00000 & 0.00000 & 1.00000\\
     &     &     &   200 & 0.33562 & -1.00000 & -3.00000 & 0.00000 & " \\
     &     &     &  2000 & 0.33562 & -1.00000 & -3.00000 & 0.00000 & " \\
     &     &     & 20000 & 0.33562 & -1.00000 & -3.00000 & 0.00000 & " \\
 0.0 & 2.5 & 1.5 &    20 & 0.00000 & -2.00000 & -2.00000 & 0.00000 & 0.00000\\
     &     &     &   200 & 0.00000 & -2.00000 & -2.00000 & 0.00000 & " \\
     &     &     &  2000 & 0.00000 & -2.00000 & -2.00000 & 0.00000 & " \\
     &     &     & 20000 & 0.00000 & -2.00000 & -2.00000 & 0.00000 & " \\
 0.5 & 2.0 & 0.0 &    20 & 0.24254 & -1.75735 & -1.75746 & 0.22586 & 0.24254\\
     &     &     &   200 & 0.24254 & -1.75746 & -1.75746 & 0.24078 & " \\
     &     &     &  2000 & 0.24254 & -1.75746 & -1.75746 & 0.24236 & " \\
     &     &     & 20000 & 0.24254 & -1.75746 & -1.75746 & 0.24252 & " \\
\end{tabular}
\end{table}

\unitlength1cm

\end{document}